\documentclass[aps,pre,twocolumn]{revtex4}
\usepackage{amsmath,bm}
\usepackage{graphicx}

\begin{document}

\title{Strong effect of weak diffusion on scalar turbulence at large scales}

\author{M. Chertkov$^a$, I. Kolokolov$^{a,b}$, and V. Lebedev$^{a,b}$}

\affiliation{$^a$Theoretical Division \& CNLS, LANL, Los Alamos, NM 87545, USA \\
$^b$ Landau Institute for Theoretical Physics, Moscow, Kosygina 2, 119334, Russia}

\begin{abstract}

Passive scalar turbulence forced steadily is characterized by the velocity correlation
scale, $L$, injection scale, $l$, and diffusive scale, $r_d$. The scales are well
separated if the diffusivity is small, $r_d\ll l,L$, and one normally says that effects
of diffusion are confined to smaller scales, $r\ll r_d$. However, if the velocity is
single scale one finds that a weak dependence of the scalar correlations on the molecular
diffusivity persists to even larger scales, e.g. $l\gg r\gg r_d$ \cite{95BCKL}. We
consider the case of $L\gg l$ and report a counter-intuitive result -- the emergence of a
new range of large scales, $L\gg r\gg l^2/r_d$, where the diffusivity shows a strong
effect on scalar correlations.

\end{abstract}

\maketitle

Studies of passive scalar advection in a random smooth flow were pioneered by Batchelor
\cite{59Bat} and Kraichnan \cite{74Kra} who considered the opposite extremes of almost
frozen and short-correlated in time random velocity gradients. The two approaches were
later extended into a unified theory describing the statistics of scalar correlations in
a general smooth flow \cite{94SS,95CFKL-a,98CFK,99BF,01FGV}. These theoretical studies
were originally motivated by interest in explaining the so-called viscous-inertial
interval of advection at the scales smaller than the viscous, Kolmogorov, scale. However,
the theory, which is nowadays often called Batchelor flow theory, also applies to many
cases of non-turbulent but chaotic smooth flows, e.g. of the type discovered recently in
polymer solutions \cite{00GS}.

Main theoretical efforts in the field were focused on the analysis of scalar correlations
within the convective range, $r_d\ll r\ll l$, i.e. at the scales smaller than the
injection scale $l$ but larger than the diffusive scale $r_d$
\cite{95CFKL-a,95BCKL,97BFL}. The range of scales above the pumping scale, even though
very nontrivial with highly intermittent correlations \cite{99BFLL}, attracted much less
attention. In this work, we continue to discuss the domain of large scales. Complementary
to our general interest in understanding multi-point correlations in turbulence, this
study was additionally motivated by our recent interest in the condensate regime of $2d$
turbulence \cite{07CCKL}, where small scale vorticity is advected passively by the large
scale, coherent part of the flow.

On a superficial level studying correlations of a fluctuating quantity upscales from the
injection scale may seem akin to many problems in equilibrium statistical mechanics, e.g.
of the type considered in the field of critical phenomena where one studies fluctuations
of an order parameter driven by thermal noise at small scales. However, the essential
difference here is that our problem is off-equilibrium. A particularly important
consequence of this fact is an intermittent, strongly non-Gaussian statistics observed
for the problem both at the scales smaller \cite{95BCKL,99BF} and larger \cite{99BFLL}
than the pumping length $l$.

In this work we extend the analysis of Ref. \cite{99BFLL} accounting for the effects of
molecular diffusivity which were ignored in \cite{99BFLL}. A surprising result of our
study is that diffusion, even though small, dominates correlation functions of the scalar
at large scales, $r\gg l^2/r_d$. This result is Batchelor flow specific and it can be
explained in dynamical, Lagrangian, terms. The collinear anomaly, established in
\cite{95BCKL} and later discussed in \cite{97BFL,99BF,99CFKV,06Ver}, states that
Lagrangian particles released along a line in a Batchelor flow stay aligned, unless weak
diffusive effects are accounted for. The anomaly reveals itself in an angular singularity
of the passive scalar multi-point correlation functions observed near the parallel
alignment of the points \cite{95BCKL,97BFL}. Translation of the collinear anomaly from
the dynamical to statistical language goes as follows. If diffusivity is neglected, a
blob of freshly injected passive scalar is deformed by a smooth flow into a strip of the
same density. The strip contributes to correlation function of the passive scalar
provided it covers all the points where the correlations are measured. Thus the strip
should have the length $r$ of the order of the separation between the points and it
should also be oriented in a way that all the points are covered. Since the flow is
chaotic and orientation of the stripe is random, the probability to cover the points is
determined by the angular size of the stripe. In incompressible flow, the blob volume is
conserved. The volume can be estimated as $l^d$, where $d$ is the space dimensionality,
thus the cross-section of the stripe can be estimated as $l^d/r$  and the angular size of
the stripe is $\sim l^d/r^d$. This results in the $\propto r^{-d}$ scaling for the $n$-th
order correlation function of the scalar, $K^{(n)}=\langle\theta_1\cdots
\theta_n\rangle$, in the collinear geometry, i.e. when the points ${\bm r}_1,\cdots,{\bm
r}_n$ lie on a straight line, and $r$ is the size of the most separated pair of points
\cite{99BFLL}. Volume preserving stretching of the scalar blob, injected at the pumping
scale $l$, should be modified when the blob size in the contracting direction reaches
$\sim r_d$, since the diffusion blocks further contraction of the blob beyond $r_d$.

\begin{figure}
 \includegraphics[width=9cm]{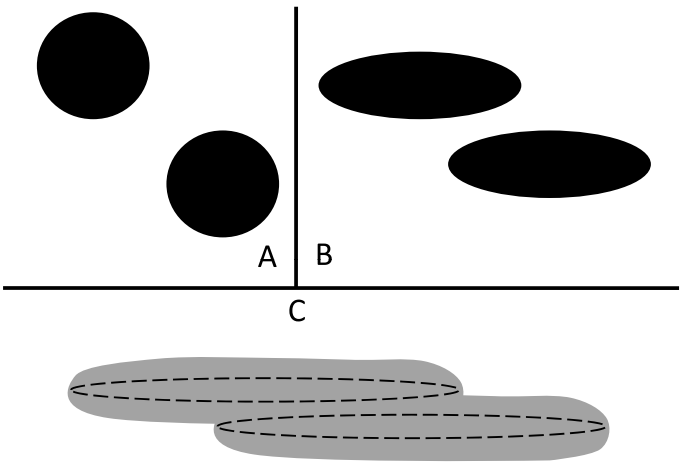}
\caption{ Figure Caption. Schematic plot illustrating Lagrangian
(temporal) evolution of two blobs of scalar. (A) Initial injection.
The blobs are of size $\sim l$ separated by the distance $\sim l$.
(B) Result of diffusion-less deformation. The blobs grow in size
along the expanding direction of the flow and decrease in size along
the contracting direction. Volumes and initial concentrations of the
scalar inside the blobs are preserved. This phase terminates when
the width of the blobs reaches $r_d$. (C) Further deformation keeps
the width of the blobs $\sim r_d$,  while the lengths of the blobs
continue to increase. Volumes of the blobs grow and the density of
the scalar inside the blobs decreases. Blobs will eventually
overlap. Dashed lines show projected shape of the blobs realized if
the diffusivity is ignored (naive extension of the stage B).}
 \label{stretch}
 \end{figure}

Let us concentrate on the $2d$ case. For a seriously stretched stripe, with the spatial
extent $r>l^2/r_d$, the stripe is $\sim r_d$-wide in cross-section and therefore
$\alpha_d\sim r_d/r$ gives an estimate for the angular size of the stripe. The temporal
dynamics of the stripe is as follows. The stripe grows in size (along the stretching
direction) while the scalar density inside the stripe, estimated as $\sim r_d/r$ fraction
of the initial density, decreases. Besides, different stripes stretched simultaneously
start to overlap along the contracting direction. Due to the random character of the
passive scalar injection, the sign of the density in the overlapping stripes alternates,
thus leading to destructive interference. This additional effect leads to further
suppression of the scalar correlations by the factor $1/\sqrt N$ where, $N\sim rr_d/l^2$,
is an estimate for the number of stripes (that were initially of size $l$ and separated
by the distance $\sim l$) which contribute to the overlapped conglomerate. Combining the
pieces, one derives the following scaling for the $2n$-th order correlation function of
the passive scalar measured at $r\gg l^2/r_d$ within the collinear geometry:
 \begin{equation}
 K^{(2n)}\propto \alpha_d N^{-n} \propto
 r^{-n-1}r_d^{1-n}.
 \label{nko}
 \end{equation}
Eq.~(\ref{nko}) is the main result of the paper, that will be confirmed below with a
proper rigor. The result shows a strong dependence of the high order correlations of the
scalar at the scales beyond the injection scales on the diffusivity, and it should thus
be contrasted with a much weaker dependence on diffusion observed in the passive scalar
correlations at the scales smaller than the injection scale \cite{95BCKL}.

Even thought Eq.~(\ref{nko}) is derived in $2d$, the qualitative
result, stating a strong sensitivity to diffusivity of the scalar
fluctuations at large scales, also extends to $3d$ (and higher
dimensions). In general, the simultaneous correlations are expressed
in terms of the Lagrangian evolution of a fluid blob, while
diffusivity stops contraction of the blobs at the diffusion scale,
$r_d$, thus making the blob globally sensitive to the small scale,
diffusion-related physics. Obviously, the $3d$ picture of the
phenomenon is more evolved due to existence of an additional, third,
dimension in the blob dynamics that can be either contracting or
expanding. As a result, the $3d$ generalization of Eq.~(\ref{nko})
becomes sensitive to the sign of the second Lyapunov exponent of the
flow. This sensitivity is similar to effects discussed in
\cite{99CFKV} and \cite{06Ver} in the contexts of kinematic dynamo,
and decaying scalar turbulence, respectively.

The dynamic equation for a passively advected scalar field, $\theta$, is
 \begin{equation}
 \partial_t{\theta}+\bm u\nabla \theta
 =\phi +\kappa\Delta\theta,
 \label{i1}
 \end{equation}
where ${\bm u}(t,{\bm r})$ is the flow velocity field, $\kappa$ is the diffusion
coefficient and $\phi(t,{\bm r})$ is the pumping term. The velocity $\bm u$ and the
forcing  $\phi$ are assumed to be independent and random functions in space/time with
prescribed statistics, spatio-temporally homogeneous and spatially isotropic. The forcing
is correlated at the scale, $l$. We consider Batchelor, that is spatially smooth, flow
where the velocity field is correlated at the scale $L$, the largest scale in the
problem. We also assume that the velocity fluctuations are sufficiently intense to
guarantee that the diffusive range, $r\ll r_d$, where the effects of advection are
strongly suppressed by diffusion, is realized at the scales smaller than the pumping
length, i.e. $r_d < l$. In Batchelor flow, the velocity difference between points
separated by a distance much smaller than $L$ is given by the first term of the Taylor
expansion in the inter-point separation, $u_{\alpha}({\bm r}_1)-u_{\alpha}({\bm r}_2)
\approx \sigma_{\alpha\beta} (r_{1;\beta}-r_{2;\beta})$. Therefore, the velocity
derivatives matrix, $\hat{\sigma}$, is the only velocity related characteristic entering
the problem at the scales smaller than $L$. In an incompressible flow, discussed here,
the velocity gradient matrix is traceless, $\mathrm{tr}\ \hat\sigma=0$. We also assume
that $\hat{\sigma}$, followed in the reference frame of a fluid parcel, is finitely
correlated in time.

Representing solution of Eq.~(\ref{i1}) in the Lagrangian frame (see \cite{98CFK} for
derivation details) one arrives at the following formal expression for the scalar field
 \begin{equation}
 \theta(t,\bm r)\!=\!\int\limits_{-\infty}^t \!\! dt'\,
 \exp\!\left(\!\kappa \!\int\limits_{t'}^t\! d\tau
 \left[\nabla \hat W(t,\tau)\right]^2\right)
 \phi\!\left(t',\bm R\right).
 \label{sca}
 \end{equation}
Here $\bm R=\hat W(t',t)\bm r$ and $\hat{W}$ is the ordered exponential
 \begin{equation}
 \hat{W}(t',t)=\mathrm{T}\exp\left(\int_{t}^{t'} d\tau\
 \hat{\sigma}(\tau)\right).
 \label{texp}
 \end{equation}
Note that in an incompressible flow, $\mathrm{det}\,\hat{W}(t)=1$.
The argument $\bm R(t')$ of the function $\phi$ in Eq. (\ref{sca})
traces back in time the Lagrangian trajectory arriving at the
position ${\bm r}$ at the moment of time $t$. The $\kappa$-dependent
exponential on the right-hand side of Eq.~(\ref{sca}) represents
effects of diffusion. Therefore, Eq. (\ref{sca}) is nothing but a
formal way to express the aforementioned qualitative arguments
concerning the Lagrangian evolution of a passive scalar blob. Since
$\phi$ is spatially correlated at the scale $l$, the temporal
integral on the right-hand side of Eq.~(\ref{sca}) is formed at
$t-t'\sim\bar\lambda^{-1} \ln(r/l)$ where $\bar\lambda$ is the
principal Lyapunov exponent of the flow, defined as the average
logarithmic rate of Lagrangian trajectories divergence. This
stretching  time diverges as $r\to\infty$, which allows us to
consider the forcing as short-correlated in time. Then the forcing
field is effectively Gaussian and thus fully described by the pair
correlation function
 \begin{equation}
 \chi\left({\bm r}-{\bm r}'\right)=
 \int dt\
 \langle \phi(t,{\bm r})\phi(t',{\bm r}')\rangle,
 \label{pum}
 \end{equation}
decaying sufficiently fast with increase in $r$ at $r>l$. We assume
that $\chi(\bm r)$ is a function of $|\bm r|$ only.

For the short-correlated forcing any correlation function of the scalar field can be
calculated in two steps. First, one averages over times larger than the pumping
correlation time but smaller than $\bar\lambda^{-1}\ln(r/l)$. This is formally equivalent
to averaging over the statistics of forcing for a given realization of the velocity
field. Averaging over velocity, corresponding to longer times and larger spatial scales,
follows. This scheme gives the following expression for the simultaneous pair correlation
function of the scalar, ${\cal K}({\bm r})=\langle\theta(t,{\bm r})\theta(t,{\bm
0})\rangle$,
 \begin{eqnarray}
 {\cal K}({\bm r})
 =\int_0^\infty\!\! dt\int d\bm k\, \left\langle
 \exp(J)\chi_{\bm k}\right\rangle,
 \label{pacod} \\
 J=i{\bm k}\hat{W}(-t,0){\bm r}
 -2{\kappa}{\bm k}\hat{I}(t){\bm k}\,,
 \label{expo} \\
 \hat{I}(t)=\int_0^{t}d\tau\,
 \hat{W}(-t,-\tau) \hat{W}^T(-t,-\tau),
 \label{imat}
 \end{eqnarray}
where $\chi_{\bm k}$ is the Fourier transform of $\chi({\bm r})$ and $T$ indicates matrix
transposition. The only averaging left to be done in Eq.~(\ref{pacod}) is over statistics
of $\hat{\sigma}$.

Consider the $d=2$ case and introduce an Iwasawa-like decomposition for the ordered
exponential
 \begin{equation}
 \hat{W}(-t,0)=
 \left(\!\begin{array}{cc}
 \cos\varphi & \sin\varphi \\
 -\sin\varphi  & \cos\varphi
 \end{array}\!\right)
 \left(\!\begin{array}{cc}
 e^{\varrho} & 0 \\
 0  & e^{-\varrho}
 \end{array}\!\right)
 \left(\!\begin{array}{cc} 1 & \zeta \\
 0  & 1  \end{array}\!\right).
 \label{ODT}
 \end{equation}
This representation is useful as the three governing fields, $\varphi,\varrho,\zeta$
decouples in the asymptotic limit of large time, $t\gg \bar{\lambda}^{-1}$. Moreover, at
the large times, the orientation angle $\varphi$ becomes random uniformly distributed
over the range $(0;2\pi)$, the distribution function of $\zeta$ freezes to a
non-universal stationary shape and the typical $\zeta$ becomes a fluctuating  $O(1)$
value, while the probability distribution of the finite time Lyapunov exponent,
$\lambda=\varrho/t$, attains the following self-similar form \cite{Ellis}
 \begin{equation}
 {\cal P}(t,\lambda)\propto
 \sqrt{t}\,\exp\left[-tS(\lambda)\right].
 \label{pl}
 \end{equation}
Here $S(\lambda)$ is the so-called Cr\'{a}mer function which is concave and achieves its
minimum at $\lambda=\bar{\lambda}$, and the $\sqrt{t}$ factor accounts for accurate
normalization of the total probability to unity. In the limit $\bar\lambda t\gg1$, the
main contribution to the integral (\ref{imat}) originates from $\tau$ close to $t$, thus
leading to
 \begin{equation}
 \label{immu}
 \hat{I}(t)= \frac{c}{\bar{\lambda}}e^{2\varrho(t)}\hat{O}(t)
 \left(\begin{array}{cc} 1 & 0 \\
 0  & 0  \end{array}\right)\, \hat{O}^{-1}(t),
 \end{equation}
where $c$ is a fluctuating factor of order unity and $\hat{O}$ is the $\varphi$-dependent
part of the decomposition (\ref{ODT}). (See \cite{98CFK} for detailed discussion of the
$c$-field statistics.) Averaging over homogeneous random orientations $\varphi$
(reflecting the assumed isotropy of the velocity fluctuations) one derives the following
expression for $J$ from Eq.~(\ref{expo})
 \begin{equation}
 J=i r\left(k_1\zeta e^{\varrho}+k_2e^{-\varrho}\right)
 -2c r_d^2 k_1^2e^{2\varrho},
 \label{pa2}
 \end{equation}
where $r_d^2=\kappa/\bar{\lambda}$ and $k_1,k_2$ are components of the wave vector $\bm
k$ in the reference frame fixed by the decomposition (\ref{ODT}).

A comparison of the two terms in Eq. (\ref{pa2}) suggests that the outer scale interval,
$r\gg l$, splits in two distinct sub-intervals: $l\ll r\ll l^2/r_d$ and $r\gg l^2/r_d$.
To describe the first interval of relatively small scales one may ignore the last term in
Eq.~(\ref{pa2}). Then, direct integration of Eq.~(\ref{pacod}) results in the
diffusionless scaling
 \begin{equation}
 {\cal K}(r)
 \propto\!\frac{1}{r^2}
 \int d^2{x}\ \chi({\bm x}),
 \label{pa4}
 \end{equation}
already derived in \cite{99BFLL}. However, evaluating the integrals in Eq.~(\ref{pacod})
in the regime where the second term in Eq.~(\ref{pa2}) dominates the first one does not
actually change the final answer for the pair correlation function (\ref{pa4}). Indeed,
in this limit, integration over $k_1$ in Eq.~(\ref{pacod}) is determined by the diffusive
exponential, that allows simply to replace $k_1$ by zero in the integrand of
Eq.~(\ref{pacod}). Integrating the result over $t$ one arrives at a factor
$\delta(k_2)/r$, while subsequent integrations over $k_2$ and over the domain of small
$\zeta$, $\zeta\ll1$, leads to the same expression for ${\cal K}(r)$, independent of the
diffusion coefficient.

We will see now that the cancelation of the $r_d$ dependence in the pair correlation
function is incidental, and it does not actually extend to the general case of higher
order correlation functions. Consider, for example, the fourth order simultaneous
correlation function, ${K}^{(4)}({\bm r}_1,{\bm r}_2,{\bm r}_3,{\bm r}_4)
=\langle\theta(t,\bm r_1)\theta(t,\bm r_2)\theta(t,\bm r_3)\theta(t,\bm r_4)\rangle$,
which is decomposed into the following sum, ${K}^{(4)}={\cal C}({\bm r}_{12},{\bm
r}_{34})+{\cal C}({\bm r}_{23},{\bm r}_{14})+ {\cal C}({\bm r}_{13},{\bm r}_{24})$, in
the case of a Gaussian pumping, where ${\bm r}_{ab}={\bm r}_a-{\bm r}_b$. Being
interested in establishing scaling law for the special case of collinear geometry, one
focuses on analysis of ${\cal C}({\bm r},{\bm r})$. Generalizing evaluations resulted in
Eqs.~(\ref{pacod},\ref{pa2}), one arrives at the following expression valid at $r\gg l$,
 \begin{eqnarray}
 {\cal C}(\bm r,\bm r)\propto
 \int_0^\infty dt_1\int_0^\infty dt_2
 \int d^2{k}\int d^2{q}
 \nonumber \\
 \biggl\langle
 \exp\Big[i r\left(k_1\zeta e^{\varrho_1}+k_2e^{-\varrho_1}
 +q_1\zeta e^{\varrho_2}+q_2e^{-\varrho_2}\right)
 \nonumber \\
 -2c r_d^2\left(k_1^2e^{2\varrho_1}+q_1^2e^{2\varrho_2}\right)
 \Big]\chi_{\bm k}\chi_{\bm q}
 \biggr\rangle \,,
 \label{fo2}
 \end{eqnarray}
where $\varrho_1=\varrho(t_1)$ and $\varrho_2=\varrho(t_2)$. If $l\ll r\ll l^2/r_d$ then
the diffusive exponent in (\ref{fo2}) can be neglected and one arrives at the
diffusionless expression
 \begin{eqnarray}
 {\cal C}(\bm r,\bm r)
 \propto\!\!\int\limits_0^\infty\!\! dt_1\,dt_2\,
 \left\langle\chi\left(r\zeta e^{\varrho_1}, r e^{-\varrho_1}  \right)
 \chi\left(r\zeta e^{\varrho_2}, r e^{-\varrho_2}\right)
 \right\rangle,
 \nonumber
 \end{eqnarray}
leading to the scaling ${\cal C}(\bm r,\bm r)\propto 1/r^2$ derived in \cite{99BFLL}.
Note that the main contribution to the above time integrals comes from the region
$\exp(\varrho_1)\sim \exp(\varrho_2)\sim r/l\gg1$. In the $r\gg l^2/r_d$ limit, the
diffusive exponential in Eq. (\ref{fo2}) cannot be replaced by unity. On the contrary, it
dominates integration over $k_1$ and $q_1$ resulting in emergence of the
$\chi_{(0,k_2)}\chi_{(0,q_2)}$ term in the integrand. Then, integrations over $t_1$ and
$t_2$ decouple from each other and one arrives at
 \begin{eqnarray}
 {\cal C}(r,r)\propto\frac{1}{r_dr^3}
 \left(\int d^2{x}\ \chi({\bm x})\right)^2,
 \label{foer}
 \end{eqnarray}
in accordance with the general formula (\ref{nko}).

The strong dependence of the correlation function on the diffusion, observed for the
collinear geometry, does not extend to a general off-collinear case,  where thus the
diffusionless consideration of \cite{99BFLL} applies. These distinct collinear and
off-collinear results are asymptotically matched in the $r_d/r$-small angular vicinity of
the collinear geometry.

Note also that if the Corrsin integral $\int d^2{r}\,\chi({\bm r})$ is equal to zero then
the leading terms (\ref{pa4}) and (\ref{foer}) in the correlation functions cancel. In
this case the behavior of the correlation functions is determined by non-universal
features of the flow statistics.

Summarizing, we have shown in this work that weak molecular diffusion does control the
large scale correlations in scalar turbulence steered by the Batchelor incompressible
flow. The main logical points of this work are: (a) Correlation of the passive scalar
within the collinear geometry are much stronger than in an off collinear case. The
angular extent of the collinear anomaly domain is controlled by the fact that a scalar
stripe injected and stretched by the flow cannot get thinner than $r_d$. (b) The effect
of diffusivity leads to a faster decay of scalar correlations with the scale $r$ at the
largest scales than in the domain of smaller scales,  where diffusion is irrelevant. (c)
Scaling in the diffusion-controlled regime becomes sensitive to the order of correlation
function and (in $3d$ or higher dimensions) on the character of the flow statistics. The
situation in $3d$, even though qualitatively similar in the sense of a general importance
of effects of diffusion, is actually more involved because of an interplay between two
distinct Lyapunov exponents characterizing the flow.

The work at Los Alamos National Laboratory was carried out under the auspices of the
National Nuclear Security Administration of the U.S. Department of Energy under Contract
No. DE-AC52-06NA25396. IK and VL also acknowledge partial support of the RFBR grant
06-02-17408-a.


\begin{thebibliography}{99}

 \bibitem{59Bat}
 G.~K.~Batchelor,
 {\it Small-scale variation of convected quantities like temperature in turbulent fluid},
 J.~Fluid.~Mech.~{\bf 5}, 113 (1959).

 \bibitem{74Kra}
 R.~H. Kraichnan,
 {\it Convection of a passive scalar by a quasi-uniform random straining field},
 J. Fluid Mech. {\bf 64}, 737 (1974).

 \bibitem{94SS}
 B.~Shraiman and E.~Siggia,
 {\it Lagrangian path integrals and fluctuations in random flow},
 Phys.~Rev.~E {\bf 49}, 2912 (1994).

 \bibitem{95CFKL-a}
 M. Chertkov, G. Falkovich, I. Kolokolov, and V. Lebedev,
 {\it Statistics of the passive scalar advected by a large-scale
 two-dimensional velocity field: analytic solution},
 Phys.~Rev.~E {\bf 51}, 5609 (1995).

 \bibitem{98CFK}
 M. Chertkov, G. Falkovich, and I. Kolokolov,
 {\it Intermittent dissipation of a passive scalar in turbulence},
 Phys. Rev. Lett. {\bf 80}, 2121 (1998).

 \bibitem{99BF}
 E. Balkovsky and A. Fouxon,
 {\it Universal Long-time Properties of Lagrangian Statistics in the
 Batchelor Regime and Their Application to the Passive Scalar Problem},
 Phys. Rev. E {\bf 60}, 4164 (1999).

 \bibitem{01FGV}
 G. Falkovich, K. Gawedzki, and M. Vergasolla,
 {\it Particles and fields in fluid turbulence},
 Rev. Mod. Phys. {\bf 73}, 913 (2001).

 \bibitem{00GS}
 {A. Groisman and V. Steinberg},
 {\it Elastic turbulence in a polymer solution flow},
 Nature {\bf 405}, 53 (2000).

 \bibitem{95BCKL}
 E. Balkovsky, M. Chertkov, I. Kolokolov, and V. Lebedev,
 {\it Fourth-order correlation function of randomly advected passive scalar},
 Pis'ma Zh. Eksp. Teor. Fiz. {\bf 61}, 1012 (1995)
 [JETP Lett. {\bf 61}, 1049 (1995)].

 \bibitem{97BFL}
 E. Balkovsky, G. Falkovich, and V. Lebedev,
 {\it Three-point correlation function of a scalar mixed
 by an almost smooth random velocity field},
 Phys. Rev. E, {\bf 55}, R4881 (1997).

 \bibitem{99BFLL}
 E. Balkovsky, G. Falkovich, V. Lebedev, and M. Lysiansky,
 {\it Large-scale properties of passive scalar advection},
 Phys. Fluids {\bf 11}, 2269 (1999)

 \bibitem{07CCKL}
 M. Chertkov, C. Connaughton, I. Kolokolov, and V. Lebedev,
 {\it Growing condensate in two-dimensional turbulence},
 submitted to Phys. Rev. Lett., nlin. CD/0612052.

 \bibitem{99CFKV}
 M. Chertkov, G. Falkovich, I. Kolokolov and M. Vergassola,
 {\it Small-scale turbulent dynamo},
 Phys. Rev. Lett. {\bf 83}, 4065 (1999).

 \bibitem{06Ver}
 S. S. Vergeles,
 {\it Spatial dependence of correlation functions in the decay
 problem for a passive scalar in a large-scale velocity field},
 ZhETF {\bf 129}, 777 (2006) [JETP, {\bf 102}, 685 (2006)].

 \bibitem{94CGK}
 M.~Chertkov, A.~Gamba, and I.~Kolokolov,
 {\it Exact field-theoretical description of passive scalar convection
 in an N-dimensional long-range velocity field},
 Phys. Lett. A {\bf 192}, 435 (1994).

 \bibitem{Ellis}
 R. Ellis, {\it Entropy, Large Deviations and Statistical Mechanics},
 Springer Verlag (1985).

\end{thebibliography}
\end{document}